\documentclass[pra,twocolumn,amsmath,amssymb,nofootinbib,superscriptaddress]{revtex4}

\usepackage{epsfig}

\begin{document}

\title{Exact Electromagnetic Casimir Energy of a Disk Opposite a Plane}

\begin{abstract}
Building on work of Meixner [J. Meixner, Z. Naturforschung {\bf 3a},
506 (1948)], we show how to compute the exact scattering
amplitude (or $T$-matrix) for electromagnetic scattering from a
perfectly conducting disk.  This calculation is a rare example of a
non-diagonal $T$-matrix  that can nonetheless be obtained in a
semi-analytic form.  We then use this result to compute the
electromagnetic Casimir interaction energy for a disk opposite a
plane, for arbitrary orientation angle of the disk, for separations
greater than the disk radius.  We find that the proximity force
approximation (PFA) significantly overestimates the Casimir energy,
both in the case of the ordinary PFA, which applies when the disk is
parallel to the plane, and the ``edge PFA,'' which applies when the
disk is perpendicular to the plane.
\end{abstract}

\author{Thorsten Emig}
\email{emig@mit.edu}
\affiliation{Laboratoire de Physique
Th\'eorique et Mod\`eles Statistiques, CNRS UMR 8626, B\^at.~100,
Universit\'e Paris-Sud, 91405 Orsay cedex, France}
\affiliation{Massachusetts Institute of Technology, MultiScale Materials Science
for Energy and Environment, Joint MIT-CNRS Laboratory (UMI 3466),
Cambridge, Massachusetts 02139, USA}
\affiliation{Massachusetts Institute of
Technology, Department of Physics, Cambridge, Massachusetts 02139, USA}

\author{Noah Graham}
\email{ngraham@middlebury.edu}
\affiliation{Department of Physics,
Middlebury College,
Middlebury, Vermont 05753, USA}

\maketitle

\section{Introduction}
\label{sec:scat-appr-casim}

Scattering methods have greatly expanded the range of situations in
which one can compute the Casimir energy \cite{Casimir48-2} of quantum
electrodynamics.  In this approach, one decomposes the path integral
representation of the Casimir energy \cite{Emig01} as a
log-determinant \cite{Kenneth06} in terms of a multiple scattering
expansion, as was done for asymptotic separations in
Ref.~\cite{Balian77,Balian78}.  This representation is closely connected
to the Krein formula \cite{Krein53,Krein62,Birman62} relating the
density of states to the scattering matrix for an ensemble of
objects. It can also be regarded as a concrete implementation of the
perspective emphasized by Schwinger \cite{Schwinger75} that the
fluctuations of the electromagnetic field can be traced back to charge
and current fluctuations on the objects.

The scattering method was first
developed for general shapes in the context of van der Waals
interactions \cite{Langbein1974}.  In planar geometries, the
scattering approach yields the Casimir energy in terms of reflection
coefficients \cite{Kats77,Jaekel91,Lambrecht06}.  By relating the
scattering matrix for a collection of spheres \cite{Henseler97} or
disks \cite{Wirzba99} to the objects' individual scattering matrices,
Bulgac, Magierski, and Wirzba were also able to use this result to
investigate the scalar and fermionic Casimir effect for disks and
spheres \cite{Bulgac01,Bulgac06,Wirzba08}.  A more general
formalism, developed in \cite{Emig07,Kenneth08,Rahi09}, has made it possible to
extend these results to other coordinate systems, an approach that is
particularly useful for geometries, such as the ones we consider here,
with edges and tips
\cite{Gies3,Gies4,wedge,parabolic1,parabolic2,Kabat1,Kabat2,Graham:2013rga,PhysRevA.91.0125012}.
It can also be applied to dilute objects in perturbation theory
\cite{Milton08-2} and extended to efficient, general-purpose numerical
calculations \cite{Reid09}; a review and further references can be found in
Ref.~\cite{Rahi2011}.
In this approach, each object is characterized by its scattering
amplitude, also known as the $T$-matrix, which describes its
response to an electromagnetic fluctuation.  It can therefore be
implemented for any object whose $T$-matrix can be calculated using a
basis for which an expansion of the free electromagnetic
Green's function exists \cite{Tai1994}.

For scalar models, the Casimir energy of a disk opposite a plane has
been calculated for a general angle between the disk axis and the
normal to the plane \cite{Emig09} as the zero-radius limit of an
oblate spheroid.  Unfortunately, for electromagnetism the wave
equation in spheroidal coordinates is not separable.  However, Meixner
\cite{Meixner} has developed a calculation of diffraction for a disk,
using a spheroidal vector basis.  By extending this calculation,
including an additional subtlety of the case where the azimuthal
quantum number $m$ is zero, we obtain the $T$-matrix in this basis
and use it to calculate the Casimir energy for a perfectly conducting
disk opposite a plane.  This $T$-matrix is nondiagonal, and the basis
in which it is expressed is not orthonormal.  Nonetheless, we can
implement appropriate conversions to make it amenable to the 
calculation of the Casimir interaction energy.  We apply this
method to the case of a disk opposite a plane, including rotations of
the disk axis relative to the normal to the plane.  This calculation
enables us to extend results for conductors with edges in Casimir
systems, giving the first example involving a compact object.

\section{The $T$-matrix}
\label{sec:t-matrix}

In this section, we calculate the $T-$matrix for an infinitely thin
and perfectly conducting disk.  Here, we build on an earlier
calculation for this scattering problem, done by Meixner in his
classic paper \cite{Meixner}.\footnote{An English translation due to
N. Sadeh is available from the authors.}  However, as we will see,
that solution was incomplete; we will extend it to obtain the full
$T$-matrix, as is required for Casimir calculations.

\subsection{Electromagnetic scattering from an infinitely thin
conducting disk}

We consider a perfectly conducting, infinitely thin disk of radius $R$
lying in the $z=0$ plane with the $z$-axis being the symmetry axis of
the disk.  This idealized case models thin disks, where the thickness
of the disk is assumed to be small compared to the wavelength of the
electromagnetic field, but large enough for the disk to be
perfectly reflecting at the wavelengths of interest.  We consider the
case of zero temperature, although it is straightforward to extend our
calculation to include thermal effects as well.

For a given incoming electric field $\mathbf{E}^{\mathrm{in}}$, we find the
corresponding outgoing wave $\mathbf{E}^{\mathrm{out}}$ such that the
boundary conditions on the disk are satisfied.  The standard boundary
conditions require that the tangential component of the electric field
$(\mathbf{E}^{\mathrm{in}}+\mathbf{E}^{\mathrm{out}})_\mathrm{tang}$
vanishes on the disk.  Were the disk a smooth body without its sharp
edge, this condition would be enough to solve the physical scattering
problem. However, the sharpness of the infinitely thin disk causes the
outgoing field to diverge on the edge.  It turns out that there are
many outgoing solutions that satisfy the boundary conditions, but
diverge at the edge in a way that the integrated electromagnetic
energy density is infinite \cite{Meixner}.  Such outgoing solutions
are nonphysical mathematical solutions of the scattering problem.
There is only one solution that diverges slowly enough such that the
electromagnetic energy density when integrated is still finite.  As a
result, this edge condition uniquely fixes the physically correct
scattering solution.

The physical scattering problem for an infinitely thin disk can then
be formulated in the following way:
\begin{enumerate}
\item The fields $(\mathbf{E}^{\mathrm{in}}$,
$\mathbf{E}^{\mathrm{out}})_\mathrm{tang}$ obey the Maxwell equations. 

\item
At large distances, the outgoing wave behaves like an outgoing spherical wave with an angular dependent amplitude.

\item
On the disk the field satisfies the boundary conditions $(\mathbf{E}^{\mathrm{in}}+\mathbf{E}^{\mathrm{out}})_\mathrm{tang}=0$.

\item
On the edge, the field satisfies the edge condition, i.e. the field
diverges slowly enough that the electromagnetic energy of the
outgoing field is finite.
\end{enumerate}

Note that the edge condition involves the outgoing field only, because
the incoming field does not diverge on the edge. Of course, the
scattering problem can equivalently be formulated in terms of the
magnetic field $\mathbf{B}$.

\subsubsection{The Debye potentials}

In the following, we use natural units where
$c=\mu_0=\varepsilon_0=1$. Following Meixner \cite{Meixner}, we
express the $\mathbf{E}$ and $\mathbf{B}$ fields in terms of scalar
Debye potentials $\Pi_1$ and $\Pi_2$,
\begin{align}
\mathbf{E}&=\nabla\times\nabla\times(\mathbf{r}\, \Pi_1)+i\, k\,
\nabla\times(\mathbf{r}\, \Pi_2), \label{3-1}
\\
\mathbf{B}&=-i\, k\, \nabla\times(\mathbf{r}\,
\Pi_1)+\nabla\times\nabla\times(\mathbf{r}\, \Pi_2).\label{3-2}
\end{align}
Here, $k$ is the wave number and $\mathbf{r}$ is the position vector
$\mathbf{r}=(x,y,z)$. The Debye potentials solve the scalar wave
equation
\begin{align}
\Delta \Pi_i+k^2\Pi_i=0, \ \ \ \text{for } i=1,2, \label{3-3}
\end{align}
and therefore the $\mathbf{E}$ and $\mathbf{B}$ fields obey the
Maxwell equations
\begin{align}
\nabla \times \mathbf{E}=i\,k\,\mathbf{B}, \ \ \ \nabla \times
\mathbf{B}=-i\,k\,\mathbf{E}. \label{3-4}
\end{align}

To express the boundary conditions
%$(\mathbf{E}^{\mathrm{in}}+\mathbf{E}^{\mathrm{out}})_\mathrm{tang}=0$
for the electric field  in terms of the Debye potentials,  it is
useful to switch to cylindrical coordinates $(\rho, \varphi, z)$. 
Due to the axial symmetry of the problem, it is sufficient to consider
Debye potentials of the form
$\Pi_{1,2}(\rho,\varphi,z)= \Pi_{1,2}(\rho,z) e^{i\, m\, \varphi}$,
where $m$ is the conserved azimuthal quantum number.
Since the incoming and the outgoing fields have the same
$\varphi$ dependence, this dependence can be
expressed as a Fourier series and considered term by term. Let us
therefore  substitute $\Pi_{1,2}(\rho,\varphi,z)= \Pi_{1,2}(\rho,z)
e^{i\, m\, \varphi}$ into Eq.~(\ref{3-1}). To eliminate the 
second derivative with respect to $z$, we use Eq.\ (\ref{3-3}). Then,
dropping the common factor of $e^{i\, m\, \varphi}$, the $\rho$ and
$\varphi$ components of the electric field $\mathbf{E}$ become
\begin{align}
E_{\rho}&=k^2 \rho \, \Pi_1(\rho,0)+2 \, \partial_{\rho} \,
\Pi_1(\rho,0)+\rho \, \partial^2_{\rho} \, \Pi_1(\rho,0), \label{3-5}
\\
E_{\varphi}&=  i\,  \frac{m \Pi_1(\rho,0)}{\rho}+ \, k \, \rho\,
\partial_z \Pi_2(\rho,0) +m \, \partial_{\rho} \,
\Pi_1(\rho,0). \label{3-6}
\end{align}
Both $E_{\rho}$ and $E_{\varphi}$ have to vanish on the disk. We first
solve Eq.\ (\ref{3-5}) for $\Pi_1(\rho,0)$ and then Eq.\ (\ref{3-6}) for
$\partial_z \Pi_2(\rho,0)$ and get
\begin{align}
\rho\,\Pi_1(\rho,0)=&\alpha \cos(k\,\rho)+\beta \, \sin(k\,\rho), \label{3-7}
\\
\rho^2\,\partial_z \Pi_2(\rho,0)=m \,(&\alpha\,
\sin(k\,\rho)-\beta\,\cos(k\,\rho)). \label{3-8}
\end{align}
Eqs.~(\ref{3-7}) and (\ref{3-8}) represent the boundary conditions
expressed in terms of the Debye potentials. The functions $\alpha$ and
$\beta$ depend on $k$ and $m$. The boundary conditions are trivially
satisfied if $\alpha=\beta=0$. Yet even the trivial solution may
violate the edge conditions if the incoming wave is not zero.  In
general, the physical solution is built out of the trivial solution
plus a special solution with nonzero $\alpha$ and $\beta$ by
exploiting the edge conditions.

Note that if $m=0$, the right-hand side of Eq.\ (\ref{3-8}) vanishes
identically. This case was not considered by Meixner in \cite{Meixner}. As
we will see, one must consider this case more carefully to avoid a free
undetermined parameter in the equations or to a situation where the
edge condition cannot be satisfied at all, resulting in an unphysical
solution. We will consider this case later on, but first we formulate
the edge conditions.

\subsubsection{The edge conditions}

Let us now use coordinates appropriate for the scattering problem.
The infinitely thin disk can be considered as a limiting case of an
oblate spheroid, so that in the following we will use oblate
spheroidal coordinates $(\xi, \eta,\varphi)$.
They are related to the Cartesian coordinates via
\begin{align}
x&=R\sqrt{(1+\xi^2)(1-\eta^2)} \, \cos(\varphi),\label{3-2-1}
\\
y&=R\sqrt{(1+\xi^2)(1-\eta^2)} \, \sin(\varphi),\label{3-2-2}
\\
z&=R\xi\eta,\label{3-2-3}
\end{align}
where
\begin{align}
0\leq\xi<\infty, \ \ \ -1\leq \eta\leq 1, \ \ \ 0\leq \varphi \leq
2\pi. \label{3-2-4}
\end{align}
The $\xi=0$ surface is then just the disk in the $z=0$ plane having
radius $R$ and the $z$-axis as a symmetry axis. The center of the disk
corresponds to $(\xi=0,\eta=\pm 1)$ and the edge is described by
$(\xi=0,\eta=0)$. We assume that the Debye potentials can be expanded
in a Taylor series in terms of $\xi$ and $\eta$ on the edge. The edge
conditions, which guarantee that the integrated energy density stays
finite, read
\cite{Meixner} 
\begin{align}
\frac{\partial \Pi_1}{\partial \xi}=\frac{\partial \Pi_1}
{\partial \eta}=\frac{\partial \Pi_2}{\partial \xi}=
\frac{\partial \Pi_2}{\partial \eta}=0 \ \ \ 
\text{for } \xi=\eta=0.\label{3-2-5}
\end{align}
To derive Eq.~(\ref{3-2-5}), we have to express $\Pi_1$ and $\Pi_2$ as a
power series in $\xi$ and $\eta$, calculate the electromagnetic field
using Eqs.~(\ref{3-1}) and (\ref{3-2}), and then integrate the
electromagnetic energy density. Then the divergences can be ruled
out by imposing Eq.~(\ref{3-2-5}).

Let us decompose the Debye potentials into incoming and outgoing parts,
\begin{align}
\Pi_i=\Pi^\mathrm{in}_i+\overline{\Pi}^\mathrm{out}_i + 
\overline{\overline{\Pi}}^\mathrm{out}_i, \ \ \ i=1,2.\label{3-2-6} 
\end{align}
Here, it useful to set
$\Pi^\mathrm{out}_i=\overline{\Pi}^\mathrm{out}_i+
\overline{\overline{\Pi}}^\mathrm{out}_i$,
$i=1,2$.
For $\overline{\Pi}^\mathrm{out}_i$ on the disk we require
\begin{align}
\Pi^\mathrm{in}_1+\overline{\Pi}^\mathrm{out}_1\equiv 0, \ \ \
\frac{\partial}{\partial z}(\Pi^\mathrm{in}_2 +
\overline{\Pi}^\mathrm{out}_2)\equiv 0 \ \ \
\text{for }\xi=0. \label{3-2-7}
\end{align}
The sum $\Pi^\mathrm{in}_i+\overline{\Pi}^\mathrm{out}_i$
represents the trivial solution in Eq.~(\ref{3-7}) and
(\ref{3-8}). The second part of the outgoing Debye potential is then
the special solution of the same Eqs.~(\ref{3-7}) and
(\ref{3-8}). Note that since the incoming wave fulfills the edge
conditions, instead of Eq.~(\ref{3-2-5}) it is sufficient to require
\begin{align}
\frac{\partial \Pi^\mathrm{out}_1}{\partial \xi}=\frac{\partial
  \Pi^\mathrm{out}_1}{\partial \eta}=\frac{\partial
  \Pi^\mathrm{out}_2}{\partial \xi}=\frac{\partial
  \Pi^\mathrm{out}_2}{\partial \eta}=0
\label{3-2-8} 
\end{align}
for $\xi=\eta=0$.
In the following sections we will derive the solution for
$\Pi_i^\mathrm{in,out}$, $\overline{\Pi}^\mathrm{out}_i$ and
$\overline{\overline{\Pi}}^\mathrm{out}_i$ in terms of spheroidal
functions.

\subsubsection{Debye potentials in terms of spheroidal functions}

There are several coordinate systems in which Eq.~(\ref{3-3}) can be
separated. For example, in spherical coordinates, every solution of
Eq.~(\ref{3-3}) can be expanded in terms of spherical waves $h_n (k r)
P_n^m(\cos \theta) \exp (i m \varphi)$, where $(r,\theta,\varphi)$
are the spherical coordinates, $n$ and $m$ are the spherical
quantum numbers, $P_n^m$ are the Legendre polynomials and $h_n$ are
the (incoming or outgoing) spherical Hankel functions. The separation
of the wave equation can also be done in spheroidal coordinates
$(\xi,\eta,\phi)$. The equivalents of the spherical radial and angular
function then are  the radial and angular spheroidal functions. The
spheroidal wave functions $L$ are called \textit{Lam\'{e}} functions
and are written as \cite{Meixner-Schaefke,Flammer}
 \begin{align}
L_{n,m}^{(1)}(\xi,\eta,\varphi;i\gamma)=S_{n,m}^{(1)}(-i\xi;i\gamma)
Sp_{n,m}(\eta;i\gamma)e^{i m \varphi},\label{3-3-1}
\\
L_{n,m}^{(3)}(\xi,\eta,\varphi;i\gamma)=S_{n,m}^{(3)}(-i\xi;i\gamma)
Sp_{n,m}(\eta;i\gamma)e^{i m \varphi},\label{3-3-2}  
\end{align}
where the first function represents the incoming wave, and the second
function the outgoing wave. In contrast to their spherical
equivalents, the radial and angular spheroidal functions, $S$ and $Sp$, 
depend on $\gamma\equiv k R$. In addition, the radial spheroidal
function also depends on $m$. Both the angular and radial spheroidal
functions become their spherical equivalents as $\gamma\rightarrow 0$
and $\xi\rightarrow \infty$, and spherical waves can be expanded in
terms of spheroidal waves and vice versa. The factors of $\pm i$ in
the arguments to the spheroidal functions correspond to the oblate case.
Finally, we note that, analogously to the spherical case,
$S_{n,m}^{(1)}(0;i\gamma)=Sp_{n,m}(0;i\gamma)=0$ for $n-m$ even and
$\partial_\xi S_{n,m}^{(1)}(0;i\gamma)=\partial_\eta
Sp_{n,m}(\eta=0;i\gamma)=0$ for $n-m$ odd. In addition
$Sp_{n,m}(\eta;i\gamma)$ is even (odd) in $\eta$ for $n-m$ even
(odd).

\subsubsection{The first part of the scattered field
$\overline{\Pi}^\mathrm{out}_i$}

Having chosen the appropriate  wave basis, let us return to the
scattering problem.
Since the Maxwell equations (\ref{3-1}) are linear in $\Pi_1$ and
$\Pi_2$, it is sufficient to restrict ourselves to the following
two cases,
\begin{align}
\Pi_1^{\mathrm{in}}=L_{n_0,m_0}^{(1)},\ \ \  \Pi_2^{\mathrm{in}}=0
\end{align}
and
\begin{align}
\Pi_1^{\mathrm{in}}=0,  \ \ \ \Pi_2^{\mathrm{in}}=L_{n_0,m_0}^{(1)}
\end{align}
for some $n_0,m_0$.  In this regard we do not consider incoming plane
waves as Meixner in \cite{Meixner}, but instead work in a basis of
vector spheroidal functions.

The first part, $\overline{\Pi}^\mathrm{out}_i$, of the decomposed
outgoing potential
$\Pi^\mathrm{out}_i=\overline{\Pi}^\mathrm{out}_i +
\overline{\overline{\Pi}}^\mathrm{out}_i$,
($i=1,2$), can then be found  straightforwardly.
Considering the  first case, $\Pi_1^{\mathrm{in}}=L_{n_0,m_0}^{(1)}$,
$\Pi_2^{\mathrm{in}}=0$, one obtains
 \begin{align}
\overline{\Pi}^\mathrm{out}_1=-L_{n_0}^{m_0\,(3)}(\xi,\eta,\varphi;i\gamma)\,
\frac{S_{n_0}^{m_0\,(1)}(-i0,i\gamma)}{S_{n_0}^{m_0\,(3)}(-i0,i\gamma)},
\ \ \ \overline{\Pi}^\mathrm{out}_2=0. \label{3-3-3}
\end{align}

For the second case $\Pi_1^{\mathrm{in}}=0$,
$\Pi_2^{\mathrm{in}}=L_{n_0,m_0}^{(1)}$, the analogous calculation
shows that
 \begin{align}
\overline{\Pi}^\mathrm{out}_1=0,  \ \
\overline{\Pi}^\mathrm{out}_2=-L_{n_0}^{m_0\,(3)}(\xi,\eta,\varphi;
i\gamma)\,\frac{S_{n_0}^{' m_0\,(1)}(-i0,i\gamma)}
{S_{n_0}^{'m_0\,(3)}(-i0,i\gamma)}. \label{3-3-4}
\end{align}
The derivative in Eq.~(\ref{3-3-4}) is taken with respect to $\xi$. To
derive Eqs.~(\ref{3-3-3}) and (\ref{3-3-4}), we used Eq.~(\ref{3-2-7}).

In general the edge conditions in Eq.~(\ref{3-2-5}) will be violated
if we substitute into Eq.~(\ref{3-2-5})
the first part $\overline{\Pi}^\mathrm{out}_i$
only. The second part $\overline{\overline{\Pi}}^\mathrm{out}_i$ is
needed to match the edge conditions. However, for some values of $n_0$
and $m_0$, the boundary conditions are satisfied by the incoming field
alone and the outgoing field vanishes identically. Since
$S_{n_0}^{m_0\,(1)}(-i0,i\gamma)\equiv 0$ for $n_0-m_0$ odd, and
$S_{n_0}^{' m_0\,(1)}(-i0,i\gamma) \equiv 0$ for $n_0-m_0$ even, there
is no scattered field for $n_0-m_0$ odd in the first case and
$n_0-m_0$ even in the second. In the next section we will construct
$\overline{\overline{\Pi}}^\mathrm{out}_i$ for general
$n_0$ and $m_0$.

\subsubsection{The second part of the scattered field
$\overline{\Pi}^\mathrm{out}_i$}

The second part of the scattered Debye potential
$\overline{\overline{\Pi}}_j^{\mathrm{sc}}$ can be expanded in terms
of outgoing waves,
\begin{align}
\overline{\overline{\Pi}}_1^{\mathrm{out}}&=
\sum_{n=|m_0|}^{\infty}A_n^{n_0,m_0}L_n^{m_0\,(3)}(\xi,\eta,\varphi,i\gamma),
\label{3-4-1}
\\
\overline{\overline{\Pi}}_2^{\mathrm{out}}&=
\sum_{n=|m_0|}^{\infty}B_n^{n_0,m_0}L_n^{m_0\,(3)}(\xi,\eta,\varphi,i\gamma)
\label{3-4-2}.
\end{align}
To get the functions $A_n^{n_0,m_0}$ and $B_n^{n_0,m_0}$,
Eqs.~(\ref{3-4-1}) and (\ref{3-4-2}) are substituted into
Eqs.~(\ref{3-7}) and (\ref{3-8}).  Using the orthogonality of the
$Sp$-functions with the normalization convention as in
\textit{Mathematica} and Meixner-Schaefke
\cite{Meixner-Schaefke},
\begin{align}\label{3-4-a}
\int_{-1}^1 Sp_n^{m}(\eta)Sp_l^{m}(\eta)\mathrm{d}\eta&=\frac{2
(n+m)!}{(2n+1)(n-m)!}\, \delta_{nl} \, ,
\end{align}
and recalling that $\rho^2=R^2(1-\eta^2)$, we can project the
expressions onto the $Sp$ functions, thus eliminating the infinite
sums.  Then $A_n^{n_0,m_0}$ and $B_n^{n_0,m_0}$ can be expressed in
terms of $\alpha$ and $\beta$ as 
\begin{align}
A_n^{n_0,m_0}&=\alpha^{n_0,m_0} a_1^{n,m_0}+\beta^{n_0,m_0}
b_1^{n,m_0},\label{3-4-3} 
\\
B_n^{n_0,m_0}&=\alpha^{n_0,m_0} a_2^{n,m_0}+\beta^{n_0,m_0}
b_2^{n,m_0}. \label{3-4-4} 
\end{align}
Here we have explicitly included the $n_0,m_0$ indices on $\alpha$
and $\beta$.  In addition, we
introduced new functions $a_1,b_1,a_2,b_2$ as 
\begin{align}
a_1^{n,m_0}&=\frac{N_{n,m_0} }{S_n^{ m_0\, (3)}(-i0) } \int_{-1}^1
Sp_n^{m_0}(\eta)\, \frac{\cos(\gamma\,\sqrt{1-\eta^2})}{R \,
  \sqrt{1-\eta^2}} \, \mathrm{d}\eta,\label{3-4-5a} 
\\
b_1^{n,m_0}&=\frac{N_{n,m_0} }{S_n^{ m_0\, (3)}(-i0) } \int_{-1}^1
Sp_n^{m_0}(\eta)\, \frac{\sin(\gamma\,\sqrt{1-\eta^2})}{R \,
  \sqrt{1-\eta^2}}\, \mathrm{d}\eta,\label{3-4-5b} 
\\
a_2^{n,m_0}&=\frac{ m_0 \, N_{n,m_0} }{S_n^{' m_0\, (3)}(-i0) }
\int_{-1}^1   Sp_n^{m_0}(\eta)\, \eta
\,\frac{\sin(\gamma\,\sqrt{1-\eta^2})}{1-\eta^2} \,
\mathrm{d}\eta,\label{3-4-5c} 
\\
b_2^{n,m_0}&=\frac{-m_0 N_{n,m_0} }{S_n^{ 'm_0\, (3)}(-i0) }
\int_{-1}^1 Sp_n^{m_0}(\eta)\, \eta
\,\frac{\cos(\gamma\,\sqrt{1-\eta^2})}{1-\eta^2} \,
\mathrm{d}\eta \label{3-4-5d} 
\end{align}
with
\begin{align}
\label{eq:Nnm}
 N_{n,m}=\frac{ (2n+1)(n-m)! }{ 2 (n+m)!} \, .
\end{align}
Note that apart from their indices, the functions  $A$, $B$, $\alpha$,
$\beta$, $a_1$, $b_1$, $a_2$ and $b_2$ depend on $\gamma$, and 
in Eqs.~(\ref{3-4-5a})--(\ref{3-4-5c}) we have also suppressed the
dependence on $\gamma$ in the functions $S$ and $Sp$.
Note also that $a_1^{n,m_0}, b_1^{n,m_0}$ vanish for $n-m_0$ odd,
and  $a_2^{n,m_0}, b_2^{n,m_0}$ vanish for $n-m_0$ even. Indeed for
$n-m_0$ odd, the function $Sp_n^{m_0}(\eta)$ is odd in $\eta$. Since
it is multiplied by an even function in $\eta$ in Eq.~(\ref{3-4-5a})
and (\ref{3-4-5b}), the integrals for $a_1$ and $b_1$
vanish. Analogously, one verifies the second case.

So far, we have strictly followed Meixner \cite{Meixner}, implicitly
assuming $m_0 \neq 0$.  For $m_0=0$, the functions $a_2$  vanishes
identically, whereas the function $b_2$ becomes ill-defined: on the
one hand, the integral in Eq.~(\ref{3-4-5d}) is multiplied by $m_0=0$,
and on the other hand, the integral itself diverges. The case $m_0=0$
therefore requires  further consideration.  For $m_0=0$,
Eq.~(\ref{3-8}) simply reads $\rho\partial_z \Pi_2=0$, meaning that
$\Pi_2$ is proportional to a $\delta$-function of $\rho$.  As a
result, we find $a_2^{n,0}=0$ and $b_2^{n,0}=-(2n+1)
(Sp_n^{0}(1)-Sp_n^{0}(-1))/2 S_n^{ 'm_0\,(3)}(-i0)$, where we write
the right hand side of Eq.~(\ref{3-4-4}) as $\tilde{\beta}^{n_0,0}
b_2^{n,0}$ for the case of $m_{0}=0$.

\subsubsection{Calculation of $\alpha^{n_0,m_0}$ and
  $\beta^{n_0,m_0}$}

In this section we calculate the functions $\alpha^{n_0,m_0}$ and
$\beta^{n_0,m_0}$ of Eq.~(\ref{3-4-3}) and (\ref{3-4-4}).  Then we
will be able to determine the second part of the scattered Debye
potential, $\overline{\overline{\Pi}}_1^{\mathrm{sc}}$, which,
together with the known first part, Eqs.~(\ref{3-3-3}) and
(\ref{3-3-4}), will eventually lead to the scattered field. For
$m_0 \neq 0$, there are two unknown functions $\alpha^{n_0,m_0}$ and
$\beta^{n_0,m_0}$, which can be calculated from the edge conditions in
Eq.~(\ref{3-2-8}).  There are four edge conditions, but it turns out
that that two of them, the second and third of Eq.~(\ref{3-2-8}), are
always fulfilled, whereas the first and the fourth yield the two equations
needed to determine $\alpha^{n_0,m_0}$ and $\beta^{n_0,m_0}$.

For $m_0=0$, as we mentioned at the end of the last section, there are
three functions, $\alpha^{n_0,m_0}$, $\beta^{n_0,m_0}$ and
$\tilde{\beta}^{n_0,0}$, that need to be determined.
At first glance, the system of three unknowns and only two equations
seems to be overdetermined.
But as we will see, it will be necessary to set $\alpha^{n_0,0}=0$,
because otherwise the scattered solution will diverge in the center of
the disk. 

Let us now consider the two following cases for incoming fields,
from which all incoming fields can be constructed.

\subsubsection{The case $\Pi^\mathrm{in}_1 \neq 0$, $\Pi^\mathrm{in}_2=0$}

Using the fourth edge condition in Eq.~(\ref{3-2-8}), we obtain
(with $\overline{\Pi}_2^{\mathrm{out}}=0$)
\begin{align}\label{4-1-1}
\partial_\eta \, \overline{\overline{\Pi}}_2^{\mathrm{out}}=0 \ \ \
\text{for } \xi=\eta=0,
\end{align}
where $\overline{\overline{\Pi}}_2^{\mathrm{out}}$ is given by
Eq.~(\ref{3-4-2}). Expressing $B_n^{n_0,m_0}$ as in Eq.~(\ref{3-4-4}),
we get
\begin{align}
\label{4-1-2}
\alpha^{n_0,m_0}\sum_{n=|m_0|+1}^{\infty}
a_2^{n,m_0}(\gamma)S_n^{m_0\, (3)}(-i0)Sp_n^{'\,m_0}(0)= \notag
\\
-\beta^{n_0,m_0}\sum_{n=|m_0|+1}^{\infty}
b_2^{n,m_0}(\gamma)S_n^{m_0\, (3)}(-i0)Sp_n^{'\,m_0}(0).
\end{align}
The sum starts at $n=|m_0|+1$, since $Sp_n^{'\,m_0}(\eta=0)=0$ for
$n-m_0$ even.  For $m_0=0$, the first series vanishes, since
$a_2^{n,0}=0$, whereas $\beta^{n_0,0}$ in Eq.~(\ref{4-1-2}) has to
be replaced by $\tilde{\beta}^{n_0,0}$. To satisfy the edge condition,
we therefore need  $\tilde{\beta}^{n_0,0}=0$, such that
$\overline{\overline{\Pi}}_2^{\mathrm{sc}}$ vanishes identically.

For $m_0\neq 0$, we can express $\beta^{n_0,m_0}$ as
\begin{align}\label{4-1-3}
\beta^{n_0,m_0}=-q_1^{m_0}(\gamma)\alpha^{n_0,m_0}.
\end{align}
The function $q_1^{m_0}(\gamma)$ can be calculated from
Eq.~(\ref{4-1-2}) as
\begin{align}\label{4-1-4}
q_1^{m_0}(\gamma)=\frac{sa_2^{m_0}(\gamma)}{sb_2^{m_0}(\gamma)},
\end{align}
where the functions $sa_2^{m_0}$ and $sb_2^{m_0}$ have been defined as
\begin{align}\label{4-1-4a}
sa_2^{m_0}(\gamma)=\sum_{n=|m_0|+1}^{\infty}a_2^{n,m_0}(\gamma)
S_n^{m_0\, (3)}(-i0)Sp_n^{'\,m_0}(0)
\end{align}
and
\begin{align}\label{4-1-4b}
sb_2^{m_0}(\gamma)=\sum_{n=|m_0|+1}^{\infty}
b_2^{n,m_0}(\gamma)S_n^{m_0\, (3)}(-i0)Sp_n^{'\,m_0}(0).
\end{align}
Note that the ratio $q_1^{m_0}(\gamma)$ does not depend on $n_0$.

Unfortunately, the series needed for calculating $q_1^{m_0}$ do not
converge if written as in Eqs.~(\ref{4-1-4a}) and (\ref{4-1-4b}).  The
reason is that the derivative with respect to $\eta$ has been put
inside the series.  However, evaluating the series with $Sp$
instead of $Sp'$ we get well behaved functions of $\eta$ with a well
defined derivative at $\eta=0$.  We will remedy this problem by
subtracting the leading term in $\gamma$, which can then be added
back in within an analytic computation.  The leading order integrals
necessary for this subtraction can be computed analytically using
Eqs.\ (\ref{A-1}), (\ref{A-2}), (\ref{A-3}), and (\ref{A-4}),
as summarized in the Appendix.

The leading order of $q_1^{m_0} (\gamma)$ can be found analytically
for any $m_0$.  For small $\gamma$ and $m_0\ge 0$ even we find
\begin{align}\label{4-1-5a}
sa_2^{m_0} (\gamma)= -\frac{\gamma (m_0-1)!!}{(m_0-2)!!} + O(\gamma^3)
\end{align}
and
\begin{align}\label{4-1-5b}
sb_2^{m_0}(\gamma) =\frac{2 m_0!!}{\pi (m_0-1)!!}+O(\gamma^2),
\end{align}
and for $m_0>0$ odd we have
\begin{align}\label{4-1-6a}
sa_2^{m_0}(\gamma) = -\frac{2 \gamma (m_0-1)!!}{\pi(m_0-2)!!}
+O(\gamma^3)
\end{align}
and
\begin{align}
sb_2^{m_0}(\gamma) = \frac{m_0!!}{(m_0-1)!!} +O(\gamma^2).
\end{align}
Subtracting the leading order from the diverging series term by term
renders them convergent and numerically evaluable. It is
straightforward to extend these results to $m_0<0$, since both sums
are invariant under $m_0\rightarrow -m_0$.

The remaining edge condition
\begin{align}\label{4-1-7}
\frac{\partial}{\partial \xi}(\overline{\Pi}_1^{\mathrm{out}} +
\overline{\overline{\Pi}}_1^{\mathrm{out}})=0 \ \ \ \text{for } \xi=\eta=0
\end{align}
fixes $\alpha^{n_0,m_0}$, which for $m_0\neq 0$ can be found from
Eqs.~(\ref{3-3-3}), (\ref{3-4-1}), (\ref{3-4-3}), and (\ref{4-1-3}),
\begin{align}\label{4-1-8}
\alpha^{n_0,m_0}_1(\gamma)=\frac{S_{n_0}^{' m_0\,(3)}(-i0)  Sp_{n_0}^{ m_0}(0)
S_{n_0}^{m_0\,(1)}(-i0)}{\left[sa_1^{m_0}(\gamma)- q_1^{m_0}
sb_1^{m_0}(\gamma)\right]S_{n_0}^{m_0\,(3)}(-i0)}.
\end{align}
Note the subscript of $\alpha^{n_0,m_0}$, which we added for clarity
since $\alpha^{n_0,m_0}$ will have a different functional form in the
second case, $\Pi^\mathrm{in}_1 = 0$, $\Pi^\mathrm{in}_2\neq 0$, to be
considered in the next section. Analogously to Eqs.~(\ref{4-1-4a}) and
(\ref{4-1-4b}), here $sa_1^{m_0}$ and $sb_1^{m_0}$ have been defined as
\begin{align}\label{4-1-9a}
sa_1^{m_0}(\gamma)=\sum_{n=|m_0|}^{\infty}a_1^{n,m_0}(\gamma)
S_n^{'m_0\, (3)}(-i0)Sp_n^{m_0}(0)
\end{align}
and
\begin{align}\label{4-1-9b}
sb_1^{m_0}(\gamma)=\sum_{n=|m_0|}^{\infty}b_1^{n,m_0}(\gamma)
S_n^{'m_0\, (3)}(-i0)Sp_n^{m_0}(0).
\end{align}

Once again, the series in Eqs.~(\ref{4-1-9a}) and (\ref{4-1-9b})
only converge if the derivative with respect to $\xi$ is taken after
the summation over $n$, so we again subtract the leading behavior at small
$\gamma$, which is responsible for the divergence.  This subtraction
can then be added back in as an analytic expression for any $m_0$. For
small $\gamma$ and $m_0 \geq 0$ even we obtain
\begin{align}\label{4-1-10}
sa_1^{m_0}(\gamma)=-\frac{(m_0-1)!!}{(m_0-2)!!} +O(\gamma^2)
\end{align}
and
\begin{align}\label{4-1-12}
sb_1^{m_0}(\gamma)=-\frac{2 m_0 !! \gamma}{\pi (m_0-1)!!}+O(\gamma^3),
\end{align}
and for $m_0 > 0$ odd we have
\begin{align}\label{4-1-11}
sa_1^{m_0}(\gamma)=-\frac{2(m_0-1)!!}{\pi(m_0-2)!!} +O(\gamma^2)
\end{align}
and
\begin{align}\label{4-1-13}
sb_1^{m_0}(\gamma)=-\frac{ m_0 !! \gamma}{(m_0-1)!!}+O(\gamma^3),
\end{align}
while for negative $m_0$ we use that these sums are odd in $m_0 \to -m_0$.

A special case arises for $m_0=0$. Eqs.~(\ref{3-4-3}) and
(\ref{3-4-4}) decouple and strictly speaking we now have to
distinguish between $\alpha, \beta$ in Eq.~(\ref{3-4-3}) and
$\alpha,\beta$ in Eq.~(\ref{3-4-4}), which are no longer related. Let
us consider Eq.~(\ref{4-1-2}) for $m_0=0$. Since $a_2^{n,0}\equiv 0$,
the left-hand side of Eq.~(\ref{4-1-2}) vanishes identically, and so
must the right-hand side.  Consequently, this implies
$\Pi_2^{\mathrm{out}}=0$. Now we are left with two unknowns,
$\alpha$ and $\beta$, in Eq.~(\ref{3-4-3}). If we keep $\alpha\neq 0$, the
derivative of the potential $\Pi_1$ with respect to $\xi$ will fail to
converge for $\eta=\pm 1$. This would imply a diverging $\mathbf{E}_n$
in the center of the disk. This divergence occurs only for $m_0=0$ and
can be cured by setting $\alpha=0$ in Eq.~(\ref{3-4-3}). Remarkably,
for $m_0>0$, the $Sp$ functions vanish at $\eta=\pm 1$ and the field
stays finite.  Thus we also luckily get rid of an overcounted
parameter.  The first term and the series over $b_1$ can then be
calculated and we find $\beta^{n_0,m_0=0}$ as a function of 
$\gamma$ by exploiting the edge condition in Eq.~(\ref{4-1-7}) to
obtain
\begin{align}\label{4-1-14}
\beta^{n_0,0}_1(\gamma)=\frac{S_{n_0}^{' 0\,(3)}(-i0)  Sp_{n_0}^{ 0}(0)
S_{n_0}^{0\,(1)}(-i0)}{sb_1^{0}(\gamma) S_{n_0}^{0\,(3)}(-i0)}.
\end{align}

\subsubsection{The case $\Pi^\mathrm{in}_1=0$, $\Pi^\mathrm{in}_2\neq 0$}

The second case $\Pi^\mathrm{in}_1=0$, $\Pi^\mathrm{in}_2\neq 0$ can
be treated in a similar way as in the previous section. Using the
first edge condition in Eq.~(\ref{3-2-5}) and noticing that
$\overline{\Pi}_1^{\mathrm{out}}=0$ [see Eq.~(\ref{3-3-4})], we obtain
\begin{align}
\label{4-2-1}
\partial_\xi\overline{\overline{\Pi}}_1^{\mathrm{out}}=0 \ \ \
\text{for } \xi=\eta=0. 
\end{align}
Expanding $\overline{\overline{\Pi}}_1^{\mathrm{out}}$ in terms of
spheroidal waves as in Eq.~(\ref{3-4-1}) and expressing
$A_n^{n_0,m_0}$ as in Eq.~(\ref{3-4-3}), we get
\begin{align}\label{4-2-2}
\alpha^{n_0,m_0} sa_1^{m_0}(\gamma)+\beta^{n_0,m_0} sb_1^{m_0}(\gamma)=0,
\end{align}
where the functions $sa_1$ and $sb_1$ are given by
Eqs.~(\ref{4-1-9a}) and (\ref{4-1-9b}).

As we explained in the previous section, for $m_0=0$ we have to set
$\alpha^{n_0,0}=0$, since otherwise $\partial_\xi\Pi_1$ will fail
to converge at $\xi=0, \ \eta=\pm 1$, leading to a diverging
electromagnetic field in the middle of the disk. Consequently
$\beta^{n_0,0}$ has to vanish in order to satisfy
Eq.~(\ref{4-2-2}), meaning that $\Pi^\mathrm{out}_1=0$.

Let us now restrict to $m_0\neq 0$ and  express $\beta^{n_0,m_0}$ as
\begin{align}\label{4-2-3}
\beta^{n_0,m_0}=-q_2^{m_0}(\gamma)\alpha^{n_0,m_0}.
\end{align}
The function $q_2^{m_0}$ can be easily calculated from Eq.~(\ref{4-2-2})
and is independent of $n_0$,
\begin{align}\label{4-2-4}
 q_2^{m_0}(\gamma)=\frac{sa_1^{m_0}(\gamma)}{sb_1^{m_0}(\gamma)}.
\end{align}
The expansion of the functions $sa_1$ and $sb_1$ for small $\gamma$ is
given in the previous section. The remaining edge condition
\begin{align}\label{4-2-5}
\partial_\eta(\overline{\Pi}_2^{\mathrm{sc}}+
\overline{\overline{\Pi}}_2^{\mathrm{sc}})=0 \ \ \ \text{for } \xi=\eta=0
\end{align}
fixes $\alpha^{n_0,m_0}$. Similarly to Eq.~(\ref{4-1-8}) we get
\begin{align}\label{4-2-6}
\alpha^{n_0,m_0}_2(\gamma)=\frac{S_{n_0}^{m_0\,(3)}(-i0)  Sp_{n_0}^{' m_0}(0)
  S_{n_0}^{' m_0\,(1)}(-i0)}{\left[sa_2^{n,m_0} - q_2^{m_0}
  sb_2^{n,m_0}\right]S_{n_0}^{' m_0\,(3)}(-i0)}.
\end{align}
Note again the subscript that we added to $\alpha^{n_0,m_0}$ in
order not to confuse the different functional forms in
Eq.~(\ref{4-1-8}) and (\ref{4-2-6}).

The case $m_0=0$ again needs special treatment. Since $a_2^{n,m_0}$
vanishes identically for $m_0=0$ [see Eq.~(\ref{3-4-5c})],
Eq.~(\ref{3-4-4}) reduces to  $B_n^{n_0,0}=\beta^{n_0,0}
b_2^{n,0}$, where we have dropped the tilde on $\beta^{n_0,0}$. We
then have
\begin{align}\label{4-2-7}
\beta^{n_0,0}_2(\gamma)=\frac{S_{n_0}^{0\,(3)}(-i0)  Sp_{n_0}^{' 0}(0)
S_{n_0}^{' 0\,(1)}(-i0)}{ sb_2^{n,0} S_{n_0}^{' 0\,(3)}(-i0)}.
\end{align}
As described above, we set
$\partial_\xi\overline{\overline{\Pi}}^{sc}_2\sim\delta(1-\eta^2)$ for
$\xi=0$. Then,
\begin{align}\label{25}
\overline{\overline{\Pi}}^{sc}_2=\beta^{n_0,0}
\sum_{n=0}^{\infty}b_2^{n,0}(\gamma)S_n^{0\, (3)}(-i\xi)Sp_n^{0}(\eta)
\end{align}
where $\beta^{n_0,0}$ is fixed by
\begin{align}\label{26}
&-S_{n_0}^{0\,(3)}(-i0)  Sp_{n_0}^{' 0}(0)
\frac{S_{n_0}^{'0\,(1)}(-i0)}{S_{n_0}^{' 0\,(3)}(-i0)} \cr
& + \beta^{n_0,0}\sum_{n=1}^{\infty}b_2^{n,0}(\gamma)
S_n^{0\, (3)}(-i0)Sp_n^{' 0}(0)=0.
\end{align}

\subsection{The T-matrix elements}

Having found the complete solution of the scattering problem, we can
express our results in terms of the $T$-matrix. The $T$-matrix depends
on the product $\gamma=k R$ and the quantum numbers $n$ and $m$. For
large distances from the disk, $k\ll 1/R$, the spheroidal
modes become spherical modes, which can be of two types:
electrical ($E$) modes (also called $TM$ modes) and magnetic
($M$) modes (also called $TE$ modes). This decomposition is a
general property of Debye potentials. The potential $\Pi_1$ alone
yields a magnetic field with vanishing radial component ($TM$ or $E$
modes) while the potential $\Pi_2$ corresponds to a vanishing radial
component of the electric field ($TE$ or $M$ modes).  Therefore,
the $T$-matrix can be split into four submatrices,
$T^\mathrm{EE}, T^\mathrm{MM}, T^\mathrm{EM}$ and $T^\mathrm{ME}$.  In
the following we show how the $T$-matrix can be constructed from the
results of the previous sections.

\subsubsection{The case $\Pi^\mathrm{in}_1\neq 0$, $\Pi^\mathrm{in}_2=0$}

As we have seen, the incoming mode $\Pi_1^{(\mathrm{in})}$ generates
outgoing fields $\Pi_1^{(\mathrm{out})}$ and
$\Pi_2^{(\mathrm{out})}$. The total potentials $\Pi_1$ and $\Pi_2$ are a
superposition of the incoming and outgoing fields and may be written
as
\begin{align}
\Pi_1 & = &&\Pi_{n_0}^{m_0\,(\mathrm{in})} &&+
&\sum_{n,m}T^\mathrm{EE}_{n,m,n_0,m_0}\Pi_n^{m\,(\mathrm{out})},\label{5-1-1}
\\
\Pi_2 & = &&0 && +
&\sum_{n,m}T^\mathrm{ME}_{n,m,n_0,m_0}\Pi_n^{m\,(\mathrm{out})}.\label{5-1-2}
\end{align}
Let us first consider the case $m_0\neq 0$. From Eqs.~(\ref{3-3-3}),
(\ref{3-4-1}) and (\ref{3-4-2}) we find
\begin{align}\label{5-1-3}
T^\mathrm{EE}_{n,m,n_0,m_0}=&-\frac{S_{n_0}^{m_0\,(1)}(-i0)}
{S_{n_0}^{m_0\,(3)}(-i0)}\delta_{n,n_0}\delta_{m,m_0}\notag \\ 
&+\alpha_1^{n_0,m_0}(a_1^{n,m_0}-q_1^{m_0} b_1^{n,m_0})\delta_{m,m_0}
\end{align}
and
\begin{align}\label{5-1-4}
T^\mathrm{ME}_{n,m,n_0,m_0}= \alpha_1^{n_0,m_0}(a_2^{n,m_0}-q_1^{m_0}
b_2^{n,m_0})\delta_{m,m_0}.
\end{align}
Note that all functions depend on $\gamma$.

For $m_0=0$, the matrix $T^\mathrm{ME}$ vanishes whereas
$T^\mathrm{EE}$ becomes
\begin{align}\label{5-1-5}
T^\mathrm{EE}_{n,m,n_0,0}=&-\frac{S_{n_0}^{0\,(1)}(-i0)}
{S_{n_0}^{0\,(3)}(-i0)}\delta_{n,n_0}\delta_{m,0}
+\beta_1^{n_0,0} b_1^{n,0}\delta_{m,0}.
\end{align}
For the Casimir interaction at large distances, it is useful to know
the behavior of the $T$-matrix at small $\gamma$. For the elements of
the  $T^{EE}$ and $T^{ME}$ matrices we find for $m_0>0$ the scaling
\begin{align}\label{5-1-6}
\frac{S_{n_0}^{ m_0\,(1)}(-i0)}{S_{n_0}^{m_0\,(3)}(-i0)}& \sim
O(\gamma^{2n_0+1}),
\\
\alpha_1^{m_0,n_0}(a_1^{n,m_0}-q_1^{m_0} b_1^{n,m_0})&\sim \cr
O(\gamma^{n_0})&(O(\gamma^{n+1})-O(\gamma^{n+3})),
\\
\alpha_1^{m_0,n_0}(a_2^{n,m_0}-q_1^{m_0} b_2^{n,m_0})&\sim \cr
O(\gamma^{n_0})&(O(\gamma^{n+2})-O(\gamma^{n+2})).
\end{align}
For non-vanishing $T^{EE}$ elements, $n_0-m_0$ and $n-m_0$ have to be
even. For non-vanishing $T^{ME}$ elements, $m_0$ has to be larger than
$0$ and $n_0-m_0$ even and $n-m_0$ odd. The matrix elements of order
$O(\gamma^3)$ are $T^{EE}_{0,0,2,0}=4\gamma^3/45i\pi$ and
$T^{EE}_{1,1,1,1}=T^{EE}_{1,-1,1,-1}=8i\gamma^3/9\pi$.

We now define the vector modes
\begin{align}\label{5-1-7}
\mathbf{M}_n^m=\nabla\times(\mathbf{r}\Pi_n^m), \ \ \
\mathbf{N}_n^m=\frac{1}{ik}\nabla\times\nabla\times(\mathbf{r}\Pi_n^m),
\end{align}
so that we can write the $\mathbf{E}$ field in the usual form
that defines the $T$-matrix,
\begin{align}\label{5-1-8}
\frac{\mathbf{E}}{i k}&=\mathbf{N}_{n_0}^{m_0\,(\mathrm{in})}
\\&+
\sum_{n,m}\left(T^{EE}_{n,m,n_0,m_0}\mathbf{N}_n^{m\,(\mathrm{out})} +
T^{ME}_{n,m,n_0,m_0}\mathbf{M}_n^{m\,(\mathrm{out})}\right), \nonumber
\end{align}
showing that our definition agrees with the one used usually for
vector spherical waves.

\subsubsection{The case $\Pi^\mathrm{in}_2\neq 0$, $\Pi^\mathrm{in}_1=0$}

The matrices $T^{MM}$ and $T^{EM}$ can be found as in the case before.
The T-matrix elements are now defined by
\begin{align}\label{5-2-1}
\Pi_2 & =&&\Pi_{n_0}^{m_0\,(\mathrm{in})} &+&&
\sum_{n,m}T^{MM}_{n,m,n_0,m_0}\Pi_n^{m\,(\mathrm{out})}
\\
\Pi_1 & =&&0 &+ &&\sum_{n,m}T^{EM}_{n,m,n_0,m_0}\Pi_n^{m\,(\mathrm{out})}.
\end{align}
We again first consider the case $m_0\neq 0$, and obtain
\begin{align}\label{5-2-2}
T^{MM}_{n,m,n_0,m_0}=&-\frac{S_{n_0}^{' m_0\,(1)}(-i0)}
{S_{n_0}^{'m_0\,(3)}(-i0)}\delta_{n,n_0}\delta_{m,m_0} \notag
\\
&+ \delta_{m,m_0}\alpha_2^{n_0,m_0}(a_2^{n,m_0}-q_2^{m_0} b_2^{n,m_0})
\end{align}
and
\begin{align}\label{5-2-3}
T^{EM}_{n,m,n_0,m_0}= \alpha_2^{n_0,m_0}(a_1^{n,m_0}-q_2^{m_0}
b_1^{n,m_0})\delta_{m,m_0}.
\end{align}

For $m_0=0$, the matrix $T^\mathrm{EM}$ vanishes,
whereas $T^\mathrm{MM}$ simplifies to
\begin{align}
\label{5-2-4}
T^\mathrm{MM}_{n,m,n_0,0}=&-\frac{S_{n_0}^{'0\,(1)}(-i0)}
{S_{n_0}^{'0\,(3)}(-i0)}\delta_{n,n_0}\delta_{m,0}
+\beta_2^{n_0,0} b_2^{n,0}\delta_{m,0}.
\end{align}
For the elements of the  $T^\mathrm{MM}$ and $T^\mathrm{EM}$ matrices we
find for $m>0$ at small $\gamma$ the scaling
\begin{align}\label{5-2-5}
\frac{S_{n_0}^{' m_0\,(1)}(-i0)}{S_{n_0}^{' m_0\,(3)}(-i0)} &\sim
O(\gamma^{2n_0+1}),
\\
\alpha_2^{m_0,n_0}(a_2^{n,m_0}-q_2^{m_0} b_2^{n,m_0})&\sim \cr
O(\gamma^{n_0+1})&(O(\gamma^{n+2})-O(\gamma^n)),
\\
\alpha_2^{m_0,n_0}(a_1^{n,m_0}-q_2^{m_0} b_1^{n,m_0})&\sim \cr
O(\gamma^{n_0+1})&(O(\gamma^{n+1})-O(\gamma^{n+1})).
\end{align}
For the non-vanishing $T^\mathrm{MM}$ elements $n_0-m_0$ and $n-m_0$
have to be odd, and for the non-vanishing $T^\mathrm{EM}$ elements $m_0$
has to be larger than $0$ and $n_0-m_0$ odd and $n-m_0$ even. The only
matrix element of $O(\gamma^3)$ is  $T^{MM}_{1,0,1,0}=4\gamma^3/9i\pi$.
(Without the contribution from the edge, we
would have obtained $T^{MM}_{1,0,1,0}=-2\gamma^3/9i\pi$.)

Finally, with the definitions of Eq.~(\ref{5-1-7}),
the $\mathbf{E}$ field can be written as
\begin{align}\label{5-2-7}
\frac{\mathbf{E}}{ik}&= \mathbf{M}_{n_0}^{m_0\,(\mathrm{in})}
\\&+\sum_{n,m}\left(T^{EM}_{n,m,n_0,m_0}\mathbf{N}_n^{m\,(\mathrm{out})}+
T^{MM}_{n,m,n_0,m_0}\mathbf{M}_n^{m\,(\mathrm{out})}\right).
\nonumber
\end{align}
which corresponds to the usual definition of T-matrix elements.

\subsection{Symmetry and unitarity of the $T$-matrix}

Because they are not eigenstates of $\hat L^2$, the modes 
in Eq.\ (\ref{5-1-7}) with the same $m$ are not orthogonal, and
so the $T$-matrix does not have the usual symmetry and unitarity
properties in this basis.  The asymmetry is particularly pronounced
for the case where $n\neq 0$ and $n_0=m=0$: these matrix elements begin
at higher order in $\gamma$ than the corresponding ones with $n=m=0$ and
$n_0 \neq 0$. This discrepancy can be traced to the behavior of the
$b_1$ coefficient in Eq.\ (\ref{3-4-5b}). Although it appears to be
$O(\gamma)$, as we will discuss below, an expansion in $\gamma$
yields an expansion of the angular spheroidal in terms of Legendre
functions; their orthogonality properties in turn lead to a cancellation
of the leading orders in $\gamma$. The true behavior is given by the
exact expression for the integral in the case where $m=0$, 
given in Eq.~(\ref{6-1-0a}), which is $O(\gamma^{2n+1})$.

As a result, it will be helpful to convert the $T$-matrix to the basis
of spherical vector waves.  There exist several normalization
conventions; we will use those of Emig et al. \cite{Rahi09}.  The vector
spherical wave functions then read, for an imaginary wave number
$k=i\kappa$ (which will be useful for the Casimir energy computation below),
\begin{align}\label{6-2-1}
\mathbf{M}_{lm}^\mathrm{(reg)}(\kappa,\mathbf{r}) & = \frac{1}{\sqrt{l(l+1)}}
\nabla\times \left(\phi_{lm}^\mathrm{(reg)}(\kappa,\mathbf{r})
\mathbf{r}\right),
\\
\mathbf{M}_{lm}^\mathrm{(out)}(\kappa,\mathbf{r}) &= \frac{1}{\sqrt{l(l+1)}}
\nabla\times \left(\phi_{lm}^\mathrm{(out)}(\kappa,\mathbf{r})
\mathbf{r}\right),
\\
\mathbf{N}_{lm}^\mathrm{(reg)}(\kappa,\mathbf{r}) & = 
\frac{1}{\kappa \sqrt{l(l+1)}} \nabla\times\nabla\times
\left(\phi_{lm}^\mathrm{(reg)}(\kappa,\mathbf{r}) \mathbf{r}\right),
\\
\mathbf{N}_{lm}^\mathrm{(out)}(\kappa,\mathbf{r}) &= 
\frac{1}{\kappa \sqrt{l(l+1)}}
\nabla\times\nabla\times \left(
\phi_{lm}^\mathrm{(out)}(\kappa,\mathbf{r}) \mathbf{r}\right),
\end{align}
where the modified spherical wave functions are
\begin{align}\label{6-2-2}
\phi_{lm}^\mathrm{(reg)}(\kappa,\mathbf{r}) =
 i_l(\kappa |\mathbf{r}|) Y_{lm}(\hat{\mathbf{r}}),
\\
\phi_{lm}^\mathrm{(out)}(\kappa,\mathbf{r}) = 
k_l(\kappa |\mathbf{r}|) Y_{lm}(\hat{\mathbf{r}}).
\end{align}
Here, $i_l(z)=\sqrt{\tfrac{\pi}{2 z}} I_{l+1/2}(z)$ is the modified
spherical Bessel function of the first kind, and
$k_l(z)=\sqrt{\tfrac{2}{\pi z}} K_{l+1/2}(z)$ is the modified spherical
Bessel function of the third kind.

It is important to note three differences between the definitions of
the spherical and spheroidal bases, one of which is nontrivial:
\begin{enumerate}
\item
The spherical basis has been written in terms of modified radial
functions,  the conventions for which introduce powers of $i$ relative
to the ordinary functions with imaginary wave number.
\item
The spherical waves have been written in terms of spherical harmonics, which
include a factor of $\sqrt{\frac{N_{l,m}}{2\pi}}$ compared
to the corresponding expression in terms of Legendre functions, the
analog of which is used in the spheroidal waves. [For the definition
of the factor $N_{l,m}$, see Eq.~(\ref{eq:Nnm})].
\item
The nontrivial difference is the normalization factor of
$\frac{1}{\sqrt{l(l+1)}}$.  Because the spheroidal functions are
not eigenstates of $\hat L^2$, no direct analog of this quantity
exists in the spheroidal case.  (The spheroidal eigenvalue plays a
similar role in separation of variables for the scalar wave equation,
but that quantity does not yield a corresponding normalization of the
vector spheroidal functions.)  It is the introduction of this quantity
in converting to spherical waves that renders the resulting basis
orthonormal.  We note also that while the spheroidal basis starts with
$n=0$, the spherical basis starts at $l=1$.
\end{enumerate}

To convert to the spherical basis, we begin from the expansion of the
spheroidal angular functions in terms of Legendre functions,
\begin{align} \label{6-1-1}
Sp_n^{m}(\eta; i\gamma)=\sum_{\nu\geq |m|}^\infty i^{\nu-n} A_{n,\nu}^m
(i\gamma) P^m_\nu(\eta),
\end{align}
where the expansion coefficients $A_{n,\nu}^m$ are obtained via recursion
relations \cite{Meixner-Schaefke}.
If $m$ is even (odd), the summation runs over even (odd) $\nu$ only, and
the coefficient $A_{n,\nu}^m$ is $O(\gamma^\nu)$ for small $\gamma$.
Note that the transformation matrix does not depend on any
coordinate. This fact can be used to obtain the transformation
formulas for spheroidal waves. At large $\xi$, the radial
spheroidal functions simplify to
\begin{align}\label{6-1-2}
S_n^{m\,(1)}(-i\xi; i\gamma)&\sim \frac{1}{\gamma \xi}
\cos\left(\gamma\xi -\frac{n+1}{2}\pi\right),
\\
S_n^{m\,(3)}(-i\xi; i\gamma)&\sim \frac{1}{\gamma \xi}
\exp\left(+i\left(\gamma\xi -\frac{n+1}{2}\pi\right)\right).
\end{align}
Multiplying Eq.~(\ref{6-1-1}) by  $S_n^{m\,(j)}(-i\xi; i\gamma) \,
e^{i m \varphi}$ yields the following transformation between scalar
waves,
\begin{align} \label{6-1-3}
L_n^{m (j)}(\xi,\eta,\varphi; i\gamma)= & \\
\sum_{\nu\geq |m|}^\infty &
A_{n,\nu}^m (i\gamma) \psi_\nu^{(j)} (kr)P^m_\nu(\cos\theta) 
e^{i m \varphi}, \nonumber
\end{align}
where $\psi_\nu$ denotes the spherical Hankel function of type $j$.
To verify Eq.~(\ref{6-1-3}), we use the asymptotic expansions of
$S_n^{m\,(j)}(-i\xi; i\gamma)$ at large $\xi$. One then immediately
realizes that Eq.~(\ref{6-1-3}) holds for large $\xi$. Since the
transformation matrix $A_{n,\nu}^m$ does not depend on any coordinate,
the relation obtained must also hold at any $\xi$.

The transformation inverse to Eq.~(\ref{6-1-3}) can be found again in the
limit of large $\xi$, in which case the radial functions can be canceled on
both sides. Expanding the $Sp$ functions as in Eq.~(\ref{6-1-1}) and using
the orthogonal relations for the Legendre polynomials similar to those
in Eq.~(\ref{3-4-a}) for spheroidal angular functions, yields
\begin{align} \label{6-1-4}
\left[A ^{-1}\right]_{\nu,n}^{m} (i\gamma)= \frac{ N_{n,m}}{N_{\nu,m}}
A_{n,\nu}^{ m} (i\gamma).
\end{align}
The inverse matrix is, as expected, the transposed matrix multiplied
by normalization factors.

Since the transformation matrix does not depend on any coordinate, the
same transformation matrix also transforms between vector waves. We
just let the operator $\nabla \times \left(\mathbf{r}\ldots\right)$
and $\nabla \times \nabla\times \left(\mathbf{r} \ldots\right)$ act on
Eq.\ (\ref{6-1-3}), passing through the matrix $A_{n,\nu}^m$.

\subsection{The $T$- Matrix in the spherical basis}

To transform to the spherical basis, we first form a rescaled $T$-matrix,
denoted by ${\cal T}$, in which each matrix element found above is
multiplied by a factor of $i^{n_0-n} \sqrt{\frac{N_{n_0,m}}{N_{n,m}}}$
to address the first two differences between the bases listed above.
This scaling makes manifest the symmetry in $m\to -m$.  We then use
the following matrix that describes the change of basis,
\begin{align}
M_{l m_l n m_n}^{P_l P_n}= &\delta_{m_l m_n}  \delta_{P_l P_n} \\
\times & \sqrt{l(l+1)}
\sqrt{\frac{N_{n,m_n}}{N_{l,m_l}}}
(-1)^{(l-n)/2} A_{l,n}^m(i\gamma) ,\nonumber
\end{align}
to convert between the spheroidal basis, indexed by $n$, $m_n$, and
polarization $P_n$, and the spherical basis, indexed by $l$, $m_l$,
and polarization $P_l$. Note that the spherical
index $l$ starts from $1$, while the spheroidal index $n$ starts from
$0$. The corresponding inverse transformation is given by
\begin{align}
\left[ M ^{-1} \right]_{n m_n l m_l}^{P_n P_l}= &
\delta_{m_n m_l} \delta_{P_n P_l} \\
\times & \frac{1} {\sqrt{l(l+1)}}
\sqrt{\frac{N_{n,m_n}}{N_{l,m_l}}} (-1)^{(l-n)/2} A_{l,n}^m(i\gamma).
\nonumber
\end{align}
We note that the prefactor $(l(l+1))^{\pm 1/2}$, which addresses
the third difference between the spheroidal and spherical bases listed
above, is implemented via the spherical index $l$, which is never zero.
We thus obtain the $T$-matrix in the spherical basis as
$\tilde{\cal T} = M {\cal T} M^{-1}$, which has the usual symmetry and
unitarity properties.

\section{The Translation Matrix and the Casimir Energy}

Having converted the T-matrix elements to the spherical basis, we are
now prepared to evaluate the Casimir energy of a disk that is parallel
to an infinite plane, using techniques
developed for the sphere-plane problem
\cite{Emig08-2,PhysRevA.78.012115,Rahi09}.  In this approach, the
Casimir energy is given as
\begin{align} \label{casimir}
E = \frac{\hbar c}{2\pi} \int_0^\infty d\kappa \ln \det 
\left(1-{\cal \tilde{\cal T}\tilde{\cal U}}\right),
\end{align}
where $\tilde {\cal T}$ is the $T$-matrix of the disk in spherical
coordinates and $\tilde{\cal U}$ combines the reflection coefficient
$r$ for the plane (see below) and the conversion matrix $D$
that expresses spherical vector waves centered at the origin of the disk in
terms of planar vector waves centered at the plane.  The matrix
elements of $\tilde{\cal U}$ are given by
\begin{align} \label{translation}
\tilde{\cal U}_{lml'm'}^{PP'} = &
\int_0^\infty\frac{k_\perp dk_\perp}{2\pi}
\frac{e^{-2d \kappa_\parallel}}{2\kappa \kappa_\parallel} \delta_{mm'} \\
\times & \sum_Q
D_{lm P,k_\perp Q} \,
r^{Q}\left(\kappa,\kappa_\parallel\right) \, 
\chi_{P'} \chi_Q
D_{l' -m' P',k_\perp Q}, \nonumber
\end{align}
where $d$ is the distance from the center of the disk to the plane,
$Q$ is the polarization of the plane wave,
$\chi_P$ is $+1$ for electric modes and $-1$ for magnetic modes,
$\kappa_\parallel = \sqrt{k_\perp^2+\kappa^2}$,
$r^{Q}\left(\kappa,\kappa_\parallel\right)$ is the Fresnel reflection
coefficient for scattering from the plane, and 
\begin{align} \label{translation-elt}
D_{lm P,k_\perp Q} = 
\sqrt{\frac{4 \pi N_{l,m}}{l(l+1)}}
\left\{\begin{array}{cc}
\displaystyle 
\frac{k_\perp}{\kappa} P_l^{m}{}'\left(\frac{\kappa_\parallel}{\kappa}\right)
& \hbox{for~} P=Q\\
\displaystyle 
\frac{im \kappa} {k_\perp} \chi_P
P_l^m\left(\frac{\kappa_\parallel}{\kappa}\right)
& \hbox{for~} P\neq Q
\end{array} \right.
\end{align}
gives the conversion between vector spherical waves and vector plane
waves in terms of the associated Legendre functions $P_l^m$ and its
derivative $P_l^{m}{}'$ with respect to its argument.
For a perfectly conducting plane,
$r^{Q}\left(\kappa,\kappa_\parallel\right)=\chi_Q = \pm 1$ for electric
and magnetic modes, respectively.

This expression is now suitable for numerical evaluation, which we
carry out in Mathematica, using routines for computing
spheroidal functions \cite{Emig09,Graham05} based on the 
package created by Falloon \cite{Falloon02}.  This code provides all
the necessary spheroidal functions, as well as the expansion
coefficients  $A_{\nu,n}^m(i\gamma)$.  Since we are carrying out this
calculation via a conversion to the spherical basis, we are restricted
to configurations with $d > R$, so that a sphere enclosing the
disk does not intersect the plane \cite{Rahi09}.  Our calculation
shows the corresponding numerical instabilities for $d < R$.

\subsection{Rotated disk}

\begin{figure}[htbp]
\includegraphics[width=0.95\linewidth]{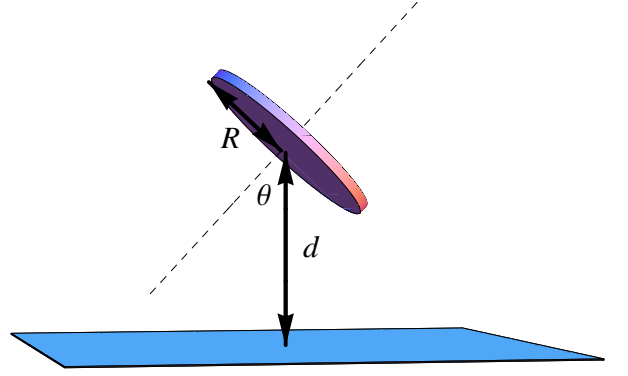}
\caption{Geometry of the disk for separation $d$, radius $R$, and
orientation angle $\theta$.}
\label{fig:diagram}
\end{figure}

The translation matrix elements in Eq.~(\ref{translation}) are
obtained from the expansion of a plane wave constructed with ``pilot
vector'' $\mathbf{\hat z}$ in terms of transverse spherical vector
modes, which are plane waves with wave vector
\begin{align}\label{k-eqn}
\mathbf{k} = & i\kappa (\sin \theta_k \cos \phi_k,\ 
\sin\theta_k \sin \phi_k,\  \cos \theta_k) \cr
= & (k_\perp \cos \phi_k,\  k_\perp \sin \phi_k,\  i \kappa_\parallel).
\end{align}
By rotating the $z$-axis of the spherical modes to an angle $\theta$
from the normal to the plane, we can obtain the Casimir energy for a
disk whose normal is tilted by that angle $\theta$ away from the
normal to the plane, allowing us to extend the results found
previously for scalar fields \cite{Emig09}.  We choose to rotate
around the $y$-axis, as shown in Fig. \ref{fig:diagram}.  In these
coordinates, the pilot vector becomes $(\sin \theta,0,\cos \theta)$, and
\begin{align}\label{angles-q}
\mathbf{k} =
i\kappa (\sin \theta_q \cos \phi_q,\ 
\sin\theta_q \sin \phi_q,\  \cos \theta_q),
\end{align}
where
\begin{align}\label{angles}
\theta_q = & \arccos \frac{i \kappa_\parallel \cos \theta
- k_\perp \cos \phi_k \sin\theta}{i\kappa},\\
\phi_q = & \arctan \frac{k_\perp \sin \phi_k}
{k_\perp \cos \phi_k \cos \theta + i \kappa_\parallel \sin \theta}.
\end{align}

The only change to the calculation is that we now have 
\begin{align} \label{translation-theta}
\tilde{\cal U}_{lml'm'}^{PP'}(\theta) = &
\int_0^\infty\frac{d^2\mathbf{k_\perp}}{(2\pi)^2}
\frac{e^{-2d \kappa_\parallel}}{2\kappa \kappa_\parallel} \\
& \times \sum_Q
D_{lm P,\mathbf{k_\perp} Q}(\theta) \,
r^{Q}\left(\kappa,\kappa_\parallel\right) \cr
& \quad \quad \chi_{P'} \chi_Q 
D_{l' -m' P',\mathbf{k_\perp} Q}(-\theta), \nonumber
\end{align}
where 
\begin{align}\label{trans-rotate-1}
D_{lm P,\mathbf{k_\perp} Q}(\theta) = & e^{im\phi_q}
\sqrt{\frac{4 \pi N_{l,m}}{l(l+1)}} \frac{\kappa}{k_\perp} \\
\times 
\Bigg[
im P_l^{m}(\cos & \theta_q) \frac{\sin \theta \sin \phi_q}{\sin \theta_q}
\cr + P_l^{m}{}'  (\cos  \theta_q) &
\sin \theta_q\Big(\cos  \theta_q \cos \phi_q \sin\theta 
- \cos \theta \sin \theta_q\Big)
\Bigg] \nonumber
\end{align}
for $P=Q$, and 
\begin{align}\label{trans-rotate-2}
D_{lm P,\mathbf{k_\perp} Q}(\theta) = & e^{im\phi_q}
\sqrt{\frac{4 \pi N_{l,m}}{l(l+1)}} \frac{\kappa}{k_\perp} \chi_P
\\
\times \Bigg[
im & P_l^{m}(\cos \theta_q) \left(\cos \theta - 
\frac{\cos \phi_q \cos \theta_q \sin \theta}{\sin \theta_q} 
\right)\cr
+ & P_l^{m}{}'(\cos \theta_q) \sin \theta \sin \theta_q \sin \phi_q
\Bigg] \nonumber
\end{align}
for $P\neq Q$.  As in the case of $\theta=0$, these expressions are
obtained as the dot product of the spherical wave and the
corresponding vector spherical harmonic of $\mathbf{\hat k}$ in the
expansion of a plane wave \cite{Varshalovich:1988ye,Forrow:2012sp}.
For any angle $\theta$, the calculation still requires $d < R$, so
that a sphere enclosing the disk does not intersect the plane. As a
result, for $\theta=\pi/2$, we could consider a disk
whose edge is arbitrarily close to the plane.  However, as the edge
approaches the plane, more partial waves and larger values of $\kappa$
are required to accurately compute the infinite sums and integrals.

We note that for $\theta \neq 0$, careful attention is
needed to avoid problems arising from branch cuts.  In particular, 
Eqs.~(\ref{trans-rotate-1}) and (\ref{trans-rotate-2})
can be expressed in terms of $k_\perp$, $\kappa_\parallel$, $\kappa$,
$\phi_k$, and $\theta$ without the need for any inverse
trigonometric functions.  Similarly, one must take care to obtain the
appropriate analytic continuation of the Legendre functions outside
the unit circle.

\subsection{Large separations}

For $d\gg R$, the Casimir energy is dominated by the contribution from
large wavelength, corresponding to small $\gamma$.  The lowest-order
contributions to the $T$-matrix are ${\cal O}(\gamma^3)$,
$T^{EE}_{0,0,2,0}=4\gamma^3/45i\pi$,
$T^{EE}_{1,1,1,1}=T^{EE}_{1,-1,1,-1}=8i\gamma^3/9\pi$, and 
$T^{MM}_{1,0,1,0}=4\gamma^3/9i\pi$.  However, $T^{EE}_{0,0,2,0}$ does
not contribute at lowest order:  Since there is no $l=0$ mode in the
spherical basis, its effect enters through off-diagonal terms mixing
different values of $l$ and $n$, which introduce additional powers
of $\gamma$.  The values of $T^{EE}_{1,1,1,1}=T^{EE}_{1,-1,1,-1}$
and $T^{MM}_{1,0,1,0}$ correspond to the static
electric and magnetic dipole responses respectively, 
$\alpha_E = 4R^3/3 \pi$ and $\alpha_M = -2R^3/3 \pi$,
which agree with previous results \cite{diskstatic}.  Using the same
approach as in the sphere-plane geometry \cite{Emig08-2}, we obtain
the Casimir energy in the long-distance limit for $\theta=0$ as
\begin{equation}
{\cal E} = -\frac{\hbar c}{8 \pi d^4} \left(2 \alpha_E - \alpha_M\right) +
{\cal O}\left(\frac{1}{d^6}\right)\,.
\label{eqn:dipole}
\end{equation}
Higher-order terms are more difficult to obtain, because they require
resummation of the infinite sums in $sa_1^{m}$,
$sb_1^{m}$, $sa_2^{m}$, and $sb_2^{m}$ at the appropriate order in $\gamma$.

\section{Results}
\label{sec:Results}

\begin{figure}[htbp]
\includegraphics[width=0.95\linewidth]{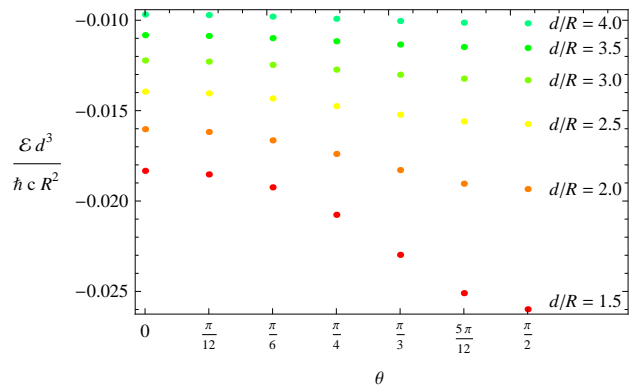}
\caption{Scaled Casimir energy for a perfectly conducting
disk of radius $R$ opposite a  perfectly conducting plane, where the
center of the disk is at a distance $d$ from the plane and the normal
to the disk is at an angle $\theta$ relative to the normal to the
plane.  The energies have been scaled by $d^3$ to facilitate comparison.}
\label{fig:rotate}
\end{figure}

Figure \ref{fig:rotate} shows the Casimir energy for a perfectly
conducting disk of radius $R$ and a perfectly conducting plane, as a
function of the rotation angle for different
separations $d/R$.  To facilitate the
comparison between difference separation distances, the energies have
been scaled by a factor of $d^3$, since a $d^{-3}$ decay is
predicted by the proximity force approximation (PFA).  The plots range
from $\theta=0$, when the disk is parallel to the plane, to
$\theta=\frac{\pi}{2}$, when the disk is perpendicular to the plane,
and from $d=1.5\ R$ to $d= 4.0\ R$. We note that at these
separations, the full energy  for $\theta=0$ is still significantly
smaller in magnitude than the prediction of the PFA, 
\begin{equation}
{\cal E}_{PFA} =  -\frac{\hbar c}{d^3} \frac{\pi^2}{720} \pi R^2,
\label{eqn:pfa}
\end{equation} 
which on this graph
would correspond to a value of $-\frac{\pi^3}{720} \approx -0.043$,
independent of $d$.  In these calculations, we have truncated the
numerical sums after $n_{\hbox{\tiny max}} = 
l_{\hbox{\tiny max}} = 5$ and used the interval 
$[\frac{1}{128 R}, \frac{4}{R}]$ for the integral over $\kappa$, and we
have checked that the results are not sensitive to these choices.
(The dimensionless ratio ${\cal E}/{\cal E}_{PFA}$ must be a function
of the dimensionless quantities $d/R$ and $\theta$.)

\begin{figure}[htbp]
\includegraphics[width=0.95\linewidth]{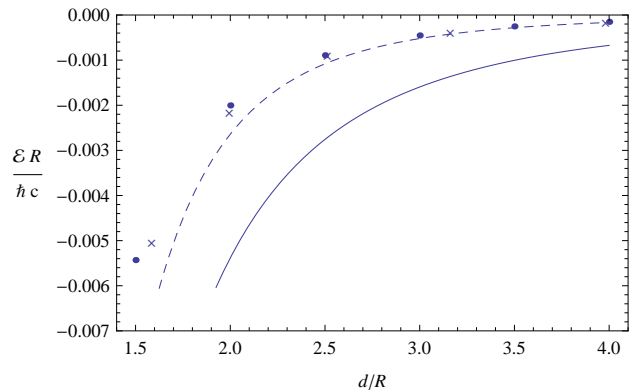}
\caption{Casimir energy for a perfectly conducting
disk of radius $R$ parallel to a perfectly conducting plane, where the
center of the disk is at a distance $d$ from the plane.  Dots
represent our results, crosses represent the 
fluctuating-surface-current calculation of Ref. \cite{Reid09},
the dotted line represents the dipole approximation, given in
Eq.~(\ref{eqn:dipole}), and the
solid line represents the PFA, given in Eq.~(\ref{eqn:pfa}).}
\label{fig:parallel}
\end{figure}

\begin{figure}[htbp]
\includegraphics[width=0.95\linewidth]{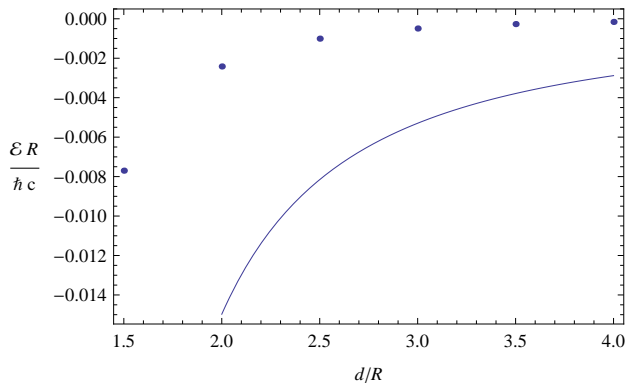}
\caption{Casimir energy for a perfectly conducting
disk of radius $R$ perpendicular to a perfectly conducting plane, where the
center of the disk is at a distance $d$ from the plane.  Dots
represent our results and the solid line represents the ``edge
PFA'' of Ref. \cite{parabolic1}.}
\label{fig:perpendicular}
\end{figure}

For the case where the disk is parallel to the plane, Fig.\
\ref{fig:parallel} shows a comparison of our result for the Casimir
energy and  the PFA prediction.  We also show numerical results
obtained by using the fluctuating-surface-current method
\cite{Reid09}.\footnote{This calculation is implemented in the
SCUFF-EM package, available from 
{\tt http://GitHub.com/HomerReid/SCUFF-EM}\,.}
The two exact methods agree well, demonstrating that the magnitude of
the energy is significantly smaller than the PFA prediction.  For the
case when the disk is perpendicular to the plane, the Casimir energy
is shown in Fig.\ \ref{fig:perpendicular}. We see that the result is
also smaller than the ``edge PFA,'' based on the result for a
half-plane with a sharp edge opposite an infinite plane \cite{parabolic1},
\begin{equation}
{\cal E}_{ePFA} = -0.0067415 \ \hbar c \pi
\sqrt{\frac{R}{2(d-R)^3}} \,.
\end{equation}

\section{Conclusions}
\label{sec:Conclusions}

Building on Meixner's analysis of diffraction from a disk
\cite{Meixner}, we have constructed the full scattering $T$-matrix for
the scattering of light from a perfectly conducting disk, which we
have then expressed in a vector spherical wave basis, a calculation
that requires particular attention to finite contributions arising 
from singular terms in the $m=0$ channel.  This result represents one
of the few cases of a non-diagonal $T$-matrix that can be computed
exactly in closed form.  The scattering approach then allows us to use
this information to obtain Casimir interaction energies for systems
such as the disk-plane geometry we have considered here, for arbitrary
orientations of the disk.  This approach is
particularly valuable for configurations where edge effects are
important, such as the case where the disk is perpendicular to the
plane, since there one cannot use a gradient expansion for gently
curved surfaces \cite{Fosco:2011xx,beyondpfa}.  We have found that the
PFA result significantly overestimates the Casimir energy at
intermediate distances, as does the ``edge PFA'' based on the result
for a half-plane.

While conversion to the vector spherical basis facilitates the
consideration of different rotation angles, it limits the calculation
to $d>R$, to ensure that a sphere enclosing the disk does not intersect the
plane.  In order to allow $d<R$, one must consider instead the vector
spheroidal basis, which is not orthonormal.  Since the scattering
method relies on a mode expansion of the free Green's function,
it cannot be applied directly to the spheroidal basis; as a result,
an important direction for future work is to generalize the scattering
method to include this case.

\acknowledgments
This work is based on preliminary studies of the scattering problem
for a disk by Alexej Weber in an earlier stage of this project. His
contribution is acknowledged.
We thank G.\ Bimonte, R.\ L.\ Jaffe, M.\ Kardar, and M.\ Kr\"uger for
helpful discussions, and M.\ T.\ H.\ Reid for carrying out
comparisons with the methods of Ref. \cite{Reid09}.
N.\ G.\ was supported in part by the National
Science Foundation (NSF) through grant PHY-1520293.

\appendix

\section{Useful Integrals}

Here we collect useful integrals, obtained from
\cite{Erdelyi,Abramowitz,Gradshteyn}.  For $m=0$, we have the
closed-form integrals
\begin{align}\label{6-1-0a}
\int_{-1}^1 Sp_n^{0}(\eta;i\gamma)\, &
\frac{\sin(\gamma\,\sqrt{1-\eta^2})}{\sqrt{1-\eta^2}}\, \mathrm{d}\eta \\
=&2 \gamma S_{n}^{0\,(1)}(0;i\gamma)^2 A_n^m(i\gamma) \nonumber
\end{align}
and
\begin{align}\label{6-1-0b}
\int_{-1}^1 Sp_n^{0}(\eta;i\gamma)\, &
\frac{\cos(\gamma\,\sqrt{1-\eta^2})}{\sqrt{1-\eta^2}}\,\mathrm{d}\eta
\\
=&2 \gamma S_{n}^{0\,(1)}(0;i\gamma) S_{n}^{0\,(2)}(0;i\gamma)
A_n^m(i\gamma) \,, \nonumber
\end{align}
where the normalization factor $A_n^m(i\gamma)$ is given by
\begin{align}\label{6-1-1-1}
A_n^m(i\gamma) =\sum_{\nu\geq |m|} i^{\nu-n} A_{n,\nu}^m (i\gamma) .
\end{align}

We can also simplify the leading-order subtractions using the integrals
\begin{align}\label{A-1}
\int_{-1}^1 P_l^m(x) dx = \frac{(-1)^l 2^{m-1} m
\Gamma\left(\frac{l}{2}\right)
\Gamma\left(\frac{l+m+1}{2}\right)}
{\Gamma\left(\frac{l+3}{2}\right) \left(\frac{l-m}{2}\right)!}
\end{align}
and
\begin{align}\label{A-2}
\int_{-1}^1 \frac{P_l^m(x)}{\sqrt{1-x^2}} dx =
\frac{2^{m} \pi (-1)^{\frac{m-l}{2}}
\Gamma\left(\frac{l+1}{2}\right)}
{\Gamma\left(\frac{1-m-l}{2}\right)
\Gamma\left(1+\frac{l}{2}\right)
\left(\frac{l-m}{2}\right)!} \,,
\end{align}
where from these results we can also obtain
\begin{align}\label{A-3}
\int_{-1}^1 \frac{x P_l^m(x)}{\sqrt{1-x^2}} dx =
\frac{2^{m} \pi (-1)^{\frac{m-l-1}{2}}
\Gamma\left(\frac{l}{2}\right)}
{\Gamma\left(-\frac{m}{2}-\frac{l}{2}\right)
\Gamma\left(\frac{l+3}{2}\right)
\Gamma\left(\frac{1-m+l}{2}\right)}
\end{align}
and
\begin{align}\label{A-4}
\int_{-1}^1 \frac{x P_l^m(x)}{1-x^2} dx =
\frac{2^{m+1} \pi (-1)^{\frac{m-l-1}{2}}
\Gamma\left(\frac{1+l}{2}\right)}
{m \, \Gamma\left(-\frac{m}{2}-\frac{l}{2}\right)
\Gamma\left(1+\frac{l}{2}\right)
\Gamma\left(\frac{1-m+l}{2}\right)}
\end{align}
using integration by parts and recurrence relations for Legendre
functions.

\bibliographystyle{apsrev}
\bibliography{article}

\end{document}